\documentclass[12pt,a4paper,aps]{article}
\usepackage{a4}
\usepackage{epsfig}
\usepackage[hmargin=2cm,vmargin=2cm,nohead]{geometry}
\begin{document}
\par
\title{The Future of Nuclear Energy: Facts and Fiction \\ 
Chapter II: \\
What is known about Secondary Uranium Resources?}

\author{
Michael Dittmar\thanks{e-mail:Michael.Dittmar@cern.ch},\\
Institute of Particle Physics,\\ 
ETH, 8093 Zurich, Switzerland \\
}
\maketitle

\begin{abstract}
During 2009 nuclear power plants, with a capacity of 370 GWe, will produce roughly 14\% of the worldwide electric energy.
About 65000 tons of natural uranium equivalent are required to operate these 
reactors. For 15 years on average only 2/3 of this fuel is provided by the uranium mines and 
1/3 comes from secondary resources. 
According to the International Atomic Energy Administration (IAEA)
and the Nuclear Energy Agency (NEA) of the Organization for Economic Co-operation and 
Development (OECD), the secondary uranium resources
will be essentially exhausted during the next 5-10 years. 
In this paper the situation concerning the secondary resources at the beginning of the year 2009 is presented.
The data used are from the 
IAEA/NEA 2007 Red Book,
``Uranium Resources, Production and Demand", and from the World Nuclear Association (WNA).

Our analysis shows that at the beginning of 2009 the remaining worldwide civilian uranium stocks 
amount to roughly 50000 tons. With the almost unavoidable yearly drawdown 
of 10000 tons, these civilian stocks will be essentially exhausted within the next 5 years. This coincides roughly with the year 2013,
when the delivery of the 10000 tons of natural uranium equivalent from russian military stocks to the USA will end.
As the majority of the remaining civilian stocks, about 30000 tons, are believed to 
be under the control of the US government and american companies, it seems rather unlikely that the 
USA will share their own  strategic uranium reserves with other large nuclear energy users.
In summary, all data indicate that a uranium supply shortage in many OECD countries can only be avoided if 
the remaining military uranium stocks from Russia and the USA, estimated to be roughly 500000 tons are made 
available to the other countries.
\end{abstract}


\newpage
\section{Introduction}
In chapter I of this analysis we have described the worldwide situation 
of nuclear energy production, the status of uranium mining and the near future perspectives and limits.
In this chapter we quantify the situation concerning secondary uranium 
resources, which have provided for the past 10-15 years the fuel for about 1/3 of the world's nuclear reactors.
The current nuclear fuel situation is,  
according to official documents from the IAEA and the NEA, totally unsustainable 
and one expects that the existing secondary resources will be exhausted within the next few years.
The seriousness of this situation, largely ignored by the media, has been expressed in 
the IAEA and NEA press declaration from the 3rd of June 2008 launching the 2007 edition of the Red Book~\cite{redbook07},\cite{redbookpress}:

{\it  ``Most secondary resources are now in decline and the gap will increasingly need to be closed by new production. Given the long lead time typically required to bring new resources into production, uranium supply shortfalls could develop if production facilities are not implemented in a timely manner.}" 

In order to clarify the importance of the secondary uranium resources 
some facts about nuclear fission energy are summarized below\cite{fission2008}:

\begin{itemize}
\item Commercial nuclear reactors are operated in 31 out of the 200 countries on our planet.
In 2009,  436 nuclear power plants, with a net installed capacity of 370.2 GW electric power, are in operation.
These reactors provide about 14\% of the worldwide produced electric energy. 
\item During the past 5-10 years, nuclear power capacity was essentially unchanged as 
the capacity increase from new reactors was compensated by the shutdown of many old reactors. 
In contrast to a claimed {\it ``nuclear renaissance"}, 
2008 was the first year since at least 40 years when not even one new reactor was connected to the electric grid.
\item    
The absolute worldwide production of electric energy from nuclear fission has, according to the WNA data base,
reached a ``peak" in 2006 when 2658 TWhe were produced.
This amount can be compared with the years 2005, 2007 and 2008 
when 2626 TWhe,  2608 TWhe and 2601 TWhe were generated
respectively\cite{fission2008}.
\item The worldwide reactor requirements for the two fissionable isotopes U235 and Pu239, 
expressed in terms of natural uranium equivalent, are currently 65000 tons, or about 170 tons/GWe, per year. 
Since more than 10 years, the primary uranium supply from the worldwide mining
provides only about 2/3 of the requirements and 1/3 comes from the drawdown of 
secondary sources\footnote{ A huge amount which corresponds almost  
to the amount extracted by the three largest uranium producing countries, Canada, Australia and Kazakhstan, together.}.
\item Out of 31 countries, which operated nuclear power plants in 2006,
only Canada, South-Africa and Russia were uranium self sufficient. The other countries use a mixture of uranium 
imports and previously accumulated uranium stocks.    
\item Nuclear power plants in Japan, South-Korea and the Western European countries, which have 
little or no uranium mining and have little or no civilian and military uranium stocks, 
are vulnerable to uranium supply shortages.  
\item About 48 reactors are under construction today and up to sixty reactors are in a discussion and planning state. 
If one assumes that all these 48 reactors can be completed in time, between 5-10 GWe/year should become operational 
during the next 5-10 years. These reactors would require roughly 500 tons of natural uranium per GWe  for the first load 
and 170 tons/year during the following years. On average 
about 5000 tons/year of uranium will thus be required for their startup and operation. 
If one assumes that the 100 older nuclear reactors are 
not terminated, the yearly uranium demand will increase from 65000 tons in 2008 to about 90000 tons by 2015. 
\end{itemize}

In the following we will analyze the status and prospects for the possible contribution 
from secondary uranium resources using the data from IAEA/NEA Red Book 2007 edition 
and from the WNA information papers. 
First we present the current composition of the secondary resources and use  
publicly available information about past uranium extraction and determine the 2009 status of uranium stocks. 
Finally we combine the information from the secondary supplies with the mining expectations and 
make a quantitative prediction for the uranium supply situation and its consequences for nuclear power plants during the next 5 years.

\section{The composition of secondary uranium resources}

As explained above, secondary uranium resources
provide the fuel for about 1/3 of the world's nuclear fission power plants.
These secondary uranium resources are classified as follows:

\begin{itemize}
\item Nuclear fuel produced from reprocessing of reactor fuels and from surplus military plutonium;
\item U235 produced by re-enrichment of previously depleted U235 uranium tails;
\item Civilian and military stocks of natural uranium and the weapon grade enriched uranium and Pu239,  
accumulated during the excess mining operation in the past 50 years.
\end{itemize}

According to the Red Book, about 3500 tons (5\% of the worldwide demand)
come from the reprocessing and from the depleted uranium tails.
An expansion of such production facilities would, like other big nuclear power projects, 
require at least 5-10 years. Such an expansion is currently not planned.

\subsection{Pu239/U235 from the reprocessing of used fuel rods}

In order to operate a standard nuclear reactor, the nuclear fuel U235 (or Pu239) has to be enriched 
to a concentration well above the concentration of 0.71\% found in natural uranium.  
New U235 enriched nuclear fuel rods contain a fraction of about 4\% of the fissionable U235 isotope and 96\% of U238. 
During the reactor operation, the U235 concentration will be reduced down to roughly 1\%. At the same time Pu239 builds up 
to an equilibrium concentration of about 1\%\footnote{The Pu239 is formed by neutron capture of U238 isotopes and subsequent 
nuclear $\beta$ decays.}. During the normal reactor cycle, the Pu239 component contributes up to 30\%  of the produced fission energy.
After a few years of operation, the fissionable material has been reduced to about 2\% and usually some new 
fuel is introduced.  Consequently, the used fuel rods still contain an interesting amount of fissionable material of U235 and Pu239.
However, nuclear fuel recycling is a rather delicate and costly operation as 
the fuel rods contain a large number of different radioactive elements. Another problem with this recycling 
is related to the military use of the Pu239 component. In the past, 
up to 95\% of the extracted Pu239 was used for military purposes, where extraction costs and associated risks were not too important.
Besides the huge costs, the potential military use of Pu239 limits the worldwide enthusiasm for nuclear fuel recycling.

In any case, some of the extracted Pu239 is used to produce the ``MOX", a mixture of 
plutonium and uranium oxides, reactor fuel\cite{MOXwiki}. Even though 
most current reactors could in principle be operated with MOX fuel,
currently only 8\% of the worldwide reactors are licensed to use this fuel.
For example, the Euratom Supply Agency (ESA) reported that, within the EU-15 countries,
the reprocessing since 1996 has produced a total of 95.8 tons Pu239. This amount corresponds to an equivalent of 11515 tons of natural uranium. The ESA reports that the year 2006 natural uranium requirements of the EU-15 reactors have been reduced with this MOX fuel  
by 1225 tons or about 5\%~\cite{euratom}.

According to the Red Book, acknowledging that 
not all countries have reported their data, the worldwide  
capacity of Pu239 recycling is about 2500 tons/year of natural uranium equivalent.

Another source for ``MOX" fuel, following an agreement in September 2000 between the 
USA and Russia,  comes from military Pu239 stocks.
Both countries agreed to convert each 34 tons at a rate of at least 2 tons per year.
During the lifetime of this agreement, this contribution adds a natural uranium equivalent of roughly 600 tons
to the secondary resources.

The used fuel rods also contain about 1\% of U235. This uranium can be partially 
recovered as re-processed uranium (or RepU). According to the 
Red Book 2007 the RepU  processing is very costly and is
currently done only by France and Russia. The yearly production capacity is estimated to up to 2500 tons, but 
only 600 tons/year are currently in operation\cite{RepU}. 
  
\subsection{U235 from depleted tails}

Depleted uranium tails are the by-product of the U235 enrichment process. The tails 
contain normally between 0.25-0.35\% of U235, or about one third of the 0.71\% in natural uranium.
The inventory of depleted uranium is increasing every year by roughly 60000 tons. It is estimated that
at the end of 2008 roughly 1 800 000 tons have been accumulated in different countries. In theory, 
a large amount of U235 is still contained in these tails, but the existing enrichment capacity is already rather limited.  
Nevertheless, during the years 2001 to 2006 Russia delivered yearly up to about 1000 tons 
of re-enriched uranium to the European Union. According to the Red Book, the Russian Federation indicated that 
this delivery will be stopped once the existing contracts end.  For the USA a pilot project is anticipated to produce a maximum 
of 1900 tons of natural uranium equivalent during a period of two years. No additional information about the status 
of this or other worldwide projects is given in the Red Book. 
 
\subsection{Past uranium extraction and how it was used} 
    
In order to understand the uranium supply situation during the coming years, 
one needs to know:
\begin{itemize}
\item How much uranium has been extracted 
in the past; 
\item how much has already been used in the reactors;
\item the geographical distribution of these stocks and 
\item how much of this excess exists in civilian and in military stockpiles. 
\end{itemize}

Some answers to these points can be obtained 
from different editions of the Red Book and from the WNA. 
Unfortunately, these presumably very precise numbers often do not agree with each other. 
For example in the Red Book 2007 edition one finds two precise, but inconsistent, numbers for the amount of extracted uranium. 
The uranium mined up to the end of 2006 is given as 2234083 tons
in chapter 1c (Table 19, page 39) and as 2325000 tons in chapter 2c (page 74), 
about 90000 tons higher.
A comparison with previous Red Book editions and the uranium mining results from 2005 and 2006 
resolves the discrepancy in favor of the higher number\footnote{Such inconsistencies in the Red Book do not give much confidence
in the claimed accuracy for many other uranium numbers.}.
 
Next, one needs to know how much of this uranium has been used (fissioned) so far.
According to the Red Book 2007 (chapter 2c) up to the end of 2006 a total of 1700 000 tons of uranium 
have been used up in reactors. Thus, the total stocks at the end of the year 2006 were 625000 tons.
During the 2007 and 2008 the world's uranium mines produced 41264 tons and 43853 tons respectively.
Another roughly 7000 tons (3500 tons/year) came from the recycling and from the reprocessing of the 
depleted uranium tails. With reactor requirements of 65000 tons/year 
one finds that the stocks have been reduced by roughly 40000 tons. 
Out of this, roughly 20000 tons came from the drawdown of the Russian military stocks
and another 20000 tons from the drawdown of the remaining civilian stocks.
Following this estimate, one finds that at the end of the year 2008 about 587000 tons of natural uranium equivalent 
remain in the military and civilian stocks.

In order to understand the supply situation from these secondary resources during the next few years 
it is important to know that the yearly delivery of 10000 tons of uranium from the Russian military stocks 
will end in 2013. 
The future of the secondary uranium supply depends thus mainly on the size of the remaining civilian uranium reserves.
Unfortunately, only a few countries have provided this information for the Red Book 2007 edition but at least 43844 tons (end of 2006)
were associated to the civilian stocks. The majority of this amount, 41279 tons, is assigned to 
the civilian stocks of the USA\cite{USAredbookp367}. 
It is further specified that roughly one half of these stocks, or 17796 tons, are owned by the US government
and that this amount is reserved to guarantee uranium supplies for their reactors and for two years.
Assuming that the yearly drawdown of civilian stocks has continued during the past two years, 
one can expect that the stocks in the USA have been reduced to an amount of 25000-30000 tons.
However, it is possible that this reduction was somewhat smaller as, unknown to the author, 
some contribution might have come from a conversion of the military stocks of the USA.

Slightly more accurate numbers can be obtained if one combines the well documented uranium 
data of the past eight years with the those presented at the 2001 annual symposium of the 
World Nuclear Association\cite{WNAsymp2001}. 
In this document, the uranium associated to the civilian and military stocks of the western and eastern blocks 
has been estimated.  The WNA analysis indicated that the civilian stocks at end of the year 2000 consisted of  
about 140000 tons, out of which 117 000 tons 
should be associated with the western block.

The WNA analysis started from a total of 1999 000 tons of extracted uranium up to the end of the year 
2000\footnote{This number is about 3\% larger than the corresponding number of 1938000 tons 
given in the Red Book 2003.}. 
The total reactor requirements up to the year 2000 were given as 1138000 tons, 
which is about 170000 tons smaller than the amount which can be calculated from the Red Book 2007 estimate.
The discrepancy between these two numbers might be understood from a different accounting 
of the remaining, not yet used, fissionable material in the reactors. 
The first uranium load requirement for a 1 GWe reactor is about 500 tons, 
but only about 170 tons are used and exchanged every year. 
Accordingly, one finds that the not yet used fuel within all existing 370 GWe reactor cores
corresponds to an equivalent of up to 185000 tons, in good agreement with the above discrepancy of 170000 tons.  
In absence of a better number, we will thus use the Red Book 2007 
number for the reactor used uranium and assume that the civilian uranium stocks at the end of the year 2000 were 
140000 tons. 

During the past eight years world uranium stocks have been reduced by about 170000 tons, or about 21000 tons/year.
While about 80000 tons came from the reduction of the Russian military stocks,  
it can be assumed that the other 90000 tons came mostly from the western civilian stocks.

It thus seems likely that at the end of 2008 only 50000 tons of civilian uranium remain; 
out of these about 27000 tons will be controlled by the USA and another 23000 tons by Russia.
If one subtracts this number from the total remaining stocks, the military stockpiles, shared somehow between the
USA and Russia, can be estimated to be roughly 540000 tons. Our estimate for the military stocks 
is at least 10\% smaller than what one would calculate from an update of the 2000 WNA estimates alone. 
If one assumes that the split between the eastern and western military stockpiles from the year 2001 WNA analysis 
were roughly correct, the military stocks at the end of 2008 can be estimated. 
Taking into account that the Russian reserves have been reduced by about 80000 tons and assuming 
the military reserves are shared mainly between the USA and Russia,  
we estimate their stocks at the beginning of 2009 to be 230000 tons and 310000 tons respectively.  

All above approximate numbers, as summarized in Table 1, 
indicate that the civilian uranium reserves at the end of 2008 consist of 
roughly 50000 tons. Furthermore, one finds that  
about 27000 tons and 23000 tons remain respectively in the western and eastern civilian stockpiles.
The military stocks can be estimated to be about a factor of 10 larger and consist of about 540000 tons of natural uranium equivalent.

\small{
\begin{table}[h]
\vspace{0.3cm}
\begin{center}
\begin{tabular}{|c|c|c|c|c|}
\hline
                    & prod. uranium                & consumed uranium               &  civilian           &  military \\
                    & total          [tons]            & incl.  [tons]                             &  stocks [tons]  &  stocks [tons] \\
\hline

end of 2006    &  2325 000                     &    1700000                              &    $\geq$ 65000             &   560 000            \\
2007/2008      &      85 200                     &      123000                              &    $\approx$ -18000     &    -20000             \\
end of 2008    &  2410 200                     &    1823000                              &    $\approx$  50000     &   540 000              \\
\hline
western stocks &                                     &        &  27000 &  230000    \\
eastern stocks  &                                    &         & 23000  &  310000   \\
\hline
\end{tabular}\vspace{0.1cm}
\caption{State of the uranium extraction and use up to the end of 2008, as estimated from the Red Book 2007 and 
the WNA numbers for the years 2007 and 2008. Roughly 3500 tons/year of natural uranium equivalent 
is estimated to come from the worldwide reprocessing and this amount is subtracted from the yearly requirements 
of 2007 and 2008. Taking into account that not all countries have reported accurate data to the Red Book and that some 
inconsistencies in the accounting exists, the civilian and military reserves contain 
perhaps an uncertainty of up to $\pm$ 10\%. The eastern and western stocks are believed to be 
controlled almost entirely by Russia and the USA respectively.} 
\end{center}
\end{table}
}

\subsection{The military uranium stockpiles}

As described in the previous section, roughly 540000 tons of natural uranium equivalent
can be associated with the military reserves of the USA and Russia. 
Not all details about these military stockpiles are public, but some numbers relevant 
for the possible conversion of these stocks into reactor fuel can nevertheless be estimated. 

Data from nuclear arms negotiations between the USA and Russia and other countries indicate that 
these two countries control currently roughly equal shares and a total of 
about 95\% of all existing nuclear weapons\cite{weapons}. For the following it seems to be sufficient to consider only  
the military stocks of these two countries. 

First we estimate how much of these 540000 tons of uranium is blocked in the remaining 20000 nuclear warheads. 
It is known that the Hiroshima bomb was made of about 64 kg of uranium, with an U235 content 
of 51 kg (enrichment of 80\%).  This corresponds roughly to the critical mass, the amount required 
to start the uncontrolled chain reaction in a sphere of uncompressed bare metal of U235.
Sophisticated methods for the uranium storage and controlled compression have reduced the critical mass by factors of 2-3.
In any case, the danger of uncontrolled explosions limits the amount of the U235 content in the warheads.
It is also known that the nuclear fission bombs of today are based on U235 or Pu239 and that fusion bombs 
are started with an explosion of U235 or Pu239. 
On average the nuclear weapons of today are estimated to 
have an explosive power at least 10 times stronger than the bomb which destroyed Hiroshima 
on August 6, 1945.

In absence of precise data, one can assume that on average each nuclear weapon contains just 
the critical mass or at least 50 kg of U235. Using this assumption, one finds that the U235 of 
one nuclear bomb corresponds on average to 7 tons of natural uranium equivalent and 
that the uranium from about 25 such bombs is sufficient to operate a 1 GWe reactor for one year.
Consequently, about 140000 tons of uranium, about 1/3 of the military stockpiles, are currently blocked 
directly in the nuclear weapons.
Another large fraction of the military stocks can be assumed to exist as highly enriched weapon grade uranium, HEU. 
In order to be used as normal nuclear fuel, this has to be downgraded to reactor ready low enrichment uranium, LEU, with an U235 
fraction of 3-4\%. 
During the past years a natural uranium equivalent of 10000 tons/year has been downgraded to reactor fuel  
and this number might be considered as being roughly equal to the currently existing capacity. 
On a time scale of 5-10 years it should be possible to increase this capacity.

Theoretically, and assuming a total nuclear disarmament, 
the military uranium stockpiles would thus be sufficient to operate the current world nuclear reactors 
for about 8 years or for about 25 years assuming the current drawdown of secondary resources.
Taking the current world real politics into account, 
such a total nuclear disarmament is unfortunately not very likely.

Nevertheless, the military stockpiles are certainly large enough and even without touching the remaining 20000 warheads, 
an extension of the current policy to convert about 10000 tons/year can be imagined.
It is however not obvious that the USA and Russia will share their strategic uranium reserves with other 
users of nuclear fission energy.
In addition, and with a longer term perspective, the downgrading of large amounts of 
previously highly enriched uranium seems to be pointless as the original enrichment process was very expensive and because 
the highly enriched uranium might eventually be needed directly to fuel future Generation IV fast breeder reactors. 

\section{Secondary uranium supply, the near future}

All existing data indicate that drawdown of the civilian inventories, practiced during the past 10 years, 
has reduced the civilian uranium stocks to roughly 50000 tons. With an expected further yearly drawdown of up to 10000 tons  
and without access to the military stocks,  the civilian western uranium stocks will be exhausted by 2013. 
Furthermore, the supply situation will become even more critical as 
the delivery of the 10000 tons of military uranium stocks from Russia to the USA will also end during 2013.  
{\bf Thus one finds, in agreement with the dramatic warning from the IAEA/NEA authorities, 
that secondary uranium supplies will essentially be terminated within a few years.}

The severity of the supply situation seems to be known and acknowledged by the uranium consulting company (UXC)~\cite{UXC}
and by uranium mining cooperations.
For example some interesting numbers about the evolution of the demand and the secondary supplies 
and the required primary uranium mining were presented 
in September 2008 at the annual WNA symposium~\cite{WNAfuture}. 
The evolution of the secondary supply side was estimated to decrease by roughly 1000 tons per year 
starting from 20029 tons in 2009 and ending with 15008 tons by 2013. For the following three years up to 2016, 
a further reduction of about 2000 tons per year is assumed (The numbers for the years 2014-2016 are 
in disagreement with the 2013 termination of the yearly delivery of 10000 tons from Russia.).
The authors of this WNA study assumed that many new reactors will start during the coming 8 years 
and they estimate that the uranium demand will increase from 65000 tons in 2008 to about 85000 tons by 2013.
Some of their uranium supply and demand estimation for the next years are repeated in Table 2.

\small{
\begin{table}[h]
\vspace{0.3cm}
\begin{center}
\begin{tabular}{|c|c|c|c|c|c|c|}
\hline
year                    & 2008    & 2009  &  2010   & 2013 &2014 & 2016 \\
\hline
Macquarie guess         &  [tons]           &    [tons]        &   [tons]           &   [tons]        &  [tons]    &  [tons]       \\
\hline
total primary   &  45145$^{*}$  & 50216  & 54879  & 70004  & 76775   & 84632 \\
total secondary  &  20015  & 20029   & 18315  & 15008 & 13000 & 9433 \\  
total demand      & 65159   & 70245  &  73194  & 88022 & 89775 & 94065 \\
\hline
total capacity [GWe]     &   365   & 379  &  382  & 409 & 427 & 477 \\
\hline
\end{tabular}\vspace{0.1cm}
\caption{Forecast for the world uranium balance prediction for the years 2008-2016 according to  
the Macquarie Research Commodities predictions presented at the 2008 WNA annual symposium~\cite{WNAfuture}.
The forecast for the 2008 primary uranium number($^{*}$) was about 1200 tons larger than the now known number of
43853 tons. The latest WNA forecast for 2009 is 49375 tons and thus also about 1000 tons smaller\cite{WNA2009pred}. 
The claimed accuracy for the forecast should raise some doubts about the underlying methodology to 
guess these numbers.}
\end{center}
\end{table}
}

As discussed above, the uranium supply might become the limiting factor 
for the near future of nuclear power production.
This demand depends, among other things, on the future of the 
aging nuclear power plants and on how fast reactors, currently under construction, can be completed.
If the primary fuel supply can not be increased as quickly as required, some 
interesting worldwide decisions about the future of nuclear power can be expected. 
For example one needs to weigh the stable operation of older nuclear power plants, which require 170 tons/GWe/year,
against the stability of the early operations for new reactors which have a first load requirement of 500 tons/GWe.
Of course, the situation will be further complicated by national and regional interests. 
It is obviously difficult to imagine that the USA government will sell their strategic uranium reserves to their economic 
competitors in Japan, China or Western Europe.   

In absence of such political insights, one can nevertheless try to guess how much uranium fuel will come from 
different sources and how many existing and new nuclear power plants can be operated with this fuel 
during the coming years.  For this forecast we use uranium supply information presented in chapter I and II of this document
and assume that the demand will be limited by the possible supply.  
This ``upper" limit guess is calculated on the basis that 170 tons/GWe/year are required to fuel an already 
operational reactor and that 500 tons/GWe(new) are are needed for the first reactor load. This forecast  
is presented in Table 3 and can be compared with the one from Table 2.
The main difference comes from the mining forecast and 
the assumption that the military component of the secondary supply from Russia will stop at the end of 2013. 
Obviously the two scenarios should be checked and corrected for the real mining results during the coming years.
Interested readers should fill Table 3 with their favorite nuclear energy scenario, requiring that it is  
consistent with their future secondary and primary uranium supply estimates. 
\small{
\begin{table}[h]
\vspace{0.3cm}
\begin{center}
\begin{tabular}{|c|c|c|c|c|c|c|}
\hline
year                    & secondary           & secondary      & primary                    & fuel for         &  fuel for new            & expected      \\
                          & civilian                  & military           & (from year-1)           & plants           &  ``operating"           & production  \\
                    & [tons]                         &  [tons]              & [tons]                       & [GWe]           & plants [GWe]           &  [TWhe]  \\

\hline
2009                   &  10000          & 10000                     & 44000                        & 370                               & 2                          & 2575   \\
2010                   &    8000          & 10000                     &  $\leq$45000             & 365                               & 2                          & 2550   \\
2011                   &    7000          & 10000                     &  $\leq$46000             & 365                               & 2                          & 2550    \\
2012                   &    7000          & 10000                     &  $\leq$47000             & 365                               & 4                          & 2550    \\
2013                   &    5000          & 10000                     & $\leq$48000              & 360                               & 4                          & 2525   \\
2014                   &    5000          &    0(?)                     & $\leq$49000               & 320                               & 0                          & 2250   \\
2015                   &    5000          &    0(?)                     & $\leq$50000               & 320                               & 0                          & 2250   \\
2016                   &    5000         &    0(?)                     & $\leq$50000                & 320                               & 0                          & 2250   \\
2017                   &    5000         &    0(?)                     & $\leq$50000                & 320                               & 0                          & 2250   \\
2018                   &    5000         &    0(?)                     & $\leq$50000                & 320                               & 0                          & 2250   \\

\hline
\end{tabular}\vspace{0.1cm}
\caption{The authors upper limit forecast covering the years 2009-2018 for the worldwide natural 
uranium equivalent primary and secondary fuel supply and its consequences for nuclear fission produced electric energy in TWhe. 
This fuel based scenario assumes that the world wide uranium mining can not be increased as estimated  
by the IAEA/NEA and WNA.  The result of this scenario will be a slow, about 1\%/year, reduction of 
nuclear produced electric energy up to 2013. The decline will become much stronger after 2013 if military 
stocks will not add at least 10000 tons/year to the fuel market.}
\end{center}
\end{table}
}

Both scenarios obviously contain some guesswork and many political and economic decisions during 
a worldwide economic crisis  
can change the near future of uranium mining and the evolution of the nuclear disarmament. 
Especially critical for uranium mining will be the situation 
in Kazakhstan where the current optimistic forecast expects that by 2013 the existing and new mines will increase
the uranium output from 8500 tons (2008) to about 18000 tons/year. An increase of similar size is also hoped to come from the mines 
in Niger, Namibia and South Africa\cite{WNAfuture}.
 
In  conclusion, uranium shortages and thus reactor shutdowns can be avoided only  
if worldwide uranium mining can be increased by roughly 10\% each year (or about 5000 tons). 
While such an increase looks rather unlikely for the next few years, the presented numbers for the required 
primary uranium in 2008 and the obtained results show a shortage of about 1200 tons  
indicating that up to 1400 tons will be missing already in 2009. This amounts corresponds roughly to the 
reduction of the uranium requirements which followed the 2007 earthquake in Japan 
with an 8 GWe nuclear capacity outage.

We expect that the uranium supply situation will become especially critically for those 
countries where a large fraction of the electric energy comes from nuclear power and 
who need to important essentially 100\% of their uranium needs. This supply problem will especially
affect OECD countries in Western Europe and Japan.  One might hope that discussions about new nuclear power plants 
will consider the warning from the NEA/IAEA press declaration about the Red Book 2007 edition 
expressed in the following paragraph:

{\it ``At the end of 2006, world uranium production (39 603 tonnes) provided about 60\% 
of world reactor requirements (66 500 tonnes) for the 435 commercial nuclear reactors in operation. 
The gap between production and requirements was made up by secondary sources drawn from government 
and commercial inventories (such as the dismantling of over 12 000 nuclear warheads and the re-enrichment of uranium tails). 
Most secondary resources are now in decline and the gap will increasingly need to be 
closed by new production. Given the long lead time typically required to bring new resources into 
production, uranium supply shortfalls could develop if production facilities are not implemented in a timely manner."}


\begin{thebibliography}{99}
\bibitem{redbook07}{The detailed numbers are extracted from the Red Book 2007 edition, 
``Uranium 2007 Resources, Production and Demand". The book is 
published every two years by the IAEA/NEA an can be found at  
OECD bookshop http://www.oecdbookshop.org/oecd/display.asp?sf1=identifiers\&st1=9789264047662.
Free online versions of some past editions can be found via ``google books."}
\bibitem{redbookpress}{Nuclear Energy Agency press declaration from 3 June 2008 about the new 
edition of the Red Book 2007 ``Uranium 2007 Resources, Production and Demand" 
at http://www.nea.fr/html/general/press/2008/2008-02.html.}
\bibitem{fission2008}{For the year 2008 status and production of nuclear electric energy see for example the WNA papers at 
http://www.world-nuclear.org/info/reactors.html; http://www.world-nuclear.org/info/inf01.html and http://www.world-nuclear.org/info/nshare.html.}
\bibitem{MOXwiki}{For some details about MOX reactor fuel and further references see http://en.wikipedia.org/wiki/MOX\_fuel.}
\bibitem{euratom} {See the EURATOM supply agency report 2006 page 24. \\
http://ec.europa.eu/euratom/ar/ar2006.pdf.}
\bibitem{RepU}{See reference \cite{redbook07} page 80.}
\bibitem{USAredbookp367}{See reference \cite{redbook07} page 367.}
\bibitem{WNAsymp2001}{See the presentation of Bernard Del Frari
``The Global Nuclear Fuel Market Supply and Demand 2001-2020"
at the 2001 WNA symposium http://www.world-nuclear.org/sym/01idx.htm.
}
\bibitem{weapons}{For an overview of the nuclear weapons and the nuclear weapon states 
see http://en.wikipedia.org/wiki/List\_of\_states\_with\_nuclear\_weapons.}
\bibitem{UXC}{For more information about the Uranium Consulting Company (UXC) see http://www.uxc.com/.}
\bibitem{WNAfuture}{See the presentation of Maximilian Layton, Macquarie Capital Securities 
``The global uranium outlook: is 2008/09 a buying opportunity?" at the 2008 WNA symposium
http://www.world-nuclear.org/sym/2008/sessions/five.html.}
\bibitem{WNA2009pred}{See the July 2009 version of http://www.world-nuclear.org/info/inf23.html.}
\end{thebibliography}
\end{document}